\documentclass[preprint,12pt,a4paper]{elsarticle}

\usepackage{amssymb}

\journal{Results in Physics}

\begin{document}

\begin{frontmatter}

\title{New high-entropy alloy superconductor Hf$_{21}$Nb$_{25}$Ti$_{15}$V$_{15}$Zr$_{24}$}

\author{Naoki Ishizu$^1$}
\address{$^1$ Department of Electrical Engineering, Faculty of Engineering, Fukuoka Institute of Technology, 3-30-1 Wajiro-higashi, Higashi-ku, Fukuoka 811-0295, Japan}

\author{Jiro Kitagawa$^1$}
\address{$^1$ Department of Electrical Engineering, Faculty of Engineering, Fukuoka Institute of Technology, 3-30-1 Wajiro-higashi, Higashi-ku, Fukuoka 811-0295, Japan}
\ead{j-kitagawa@fit.ac.jp}

\begin{abstract}
High-entropy alloys are a new class of alloys, and attract much attention due to their unique properties. Since the discovery of a superconducting high-entropy alloy (HEA) in 2014, the materials research on HEA superconductor is a hot topic. We have found that Hf$_{21}$Nb$_{25}$Ti$_{15}$V$_{15}$Zr$_{24}$ body-centered-cubic (BCC) HEA is a new superconductor with the superconducting critical temperature $T_\mathrm{c}$ of 5.3 K. We briefly discussed the comparison of cocktail effect of $T_\mathrm{c}$ among BCC HEA superconductors.
\end{abstract}


\end{frontmatter}

\clearpage

\section{Introduction}
There are two well accepted definitions\cite{Gao:book} for high-entropy alloys: alloys containing at least five elements with the atomic percentage of $i$-th element $X_{i}$  5\% $\leq$ $X_{i}$ $\leq$ 35\%, and alloys with the mixing entropy $S_\mathrm{mix}$ larger than 1.5$R$, where $R$ is the gas constant.
$S_\mathrm{mix}$ for a $n$ elements alloy is given by $-R\sum^{n}_{i=1}x_{i}lnx_{i}$, where $x_{i}$ is the mole fraction of $i$-th component. 

A high-entropy alloy (HEA) generally forms a simple crystalline structure like face-centered-cubic, body-centered-cubic (BCC) or close-packed hexagonal.
Focusing on the superconducting HEAs, BCC materials are well studied, however, there are only several reports\cite{Kozelj:PRL2014,Rohr:PNAS2016,Marik:JALCOM2018}.
This fact motivated us to carry out the materials research on HEA superconductors.
Recently equimolar BCC HfNbTiVZr is reported\cite{Pacheco:InorgChem2019}, however, a secondary phase appears during annealing.
We have searched a BCC Hf-Nb-Ti-V-Zr HEA, which is stable even after annealing, by changing the atomic composition, and checked the possible superconductivity. 
In this article, we report a thus found HEA superconductor Hf$_{21}$Nb$_{25}$Ti$_{15}$V$_{15}$Zr$_{24}$.

\section{Materials and methods}
The polycrystalline sample Hf$_{21}$Nb$_{25}$Ti$_{15}$V$_{15}$Zr$_{24}$ was made by a home-made arc furnace.
The constituent elements of Hf (99.9\%), Nb (99.9\%), Ti (99.5\%), V (99.9\%) and Zr (99\%) with the stoichiometric composition were placed on a water-cooled Cu hearth and arc-melted in an Ar atmosphere.
The annealing condition was 800 $^{\circ}$C for 4 days in an evacuated quartz tube.
The measurement methods of sample characterization and temperature dependences of ac magnetic susceptibility $\chi_{ac}$ (T) and electrical resistivity $\rho$ (T) are described in the previous papers of our team\cite{Kitagawa:JMMM2018,Hamamoto:MRX2018}.

\begin{figure}
\begin{center}
\includegraphics[width=0.8\linewidth]{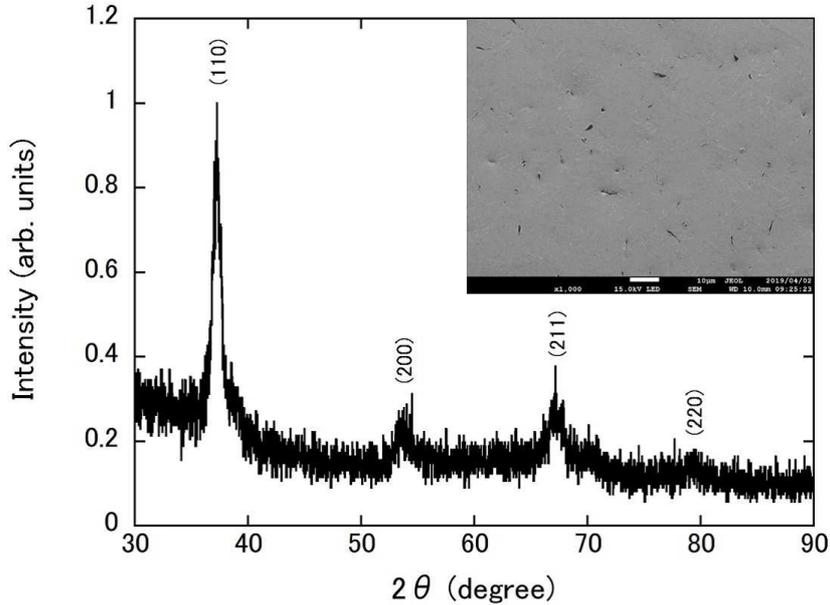}
\end{center}
\caption{X-ray diffraction pattern of Hf$_{21}$Nb$_{25}$Ti$_{15}$V$_{15}$Zr$_{24}$. The inset is the back-scattered electron (15 keV) image of the sample.}
\label{f1}
\end{figure}

\section{Results and discussion}
Figure 1 shows the X-ray diffraction pattern of Hf$_{21}$Nb$_{25}$Ti$_{15}$V$_{15}$Zr$_{24}$.
The diffraction peaks can be well indexed by a BCC structure with the lattice parameter $a$=3.401 \AA.
Back-scattered electron image obtained by a field emission scanning electron microscope with electron beams of 15 keV is shown in the inset of Fig.\ 1.
Although there are voids with small black areas, no obvious impurity phases are detected. 
The atomic composition obtained by an energy dispersive X-ray measurement is Hf$_{21.2(3)}$Nb$_{25.3(3)}$Ti$_{13.8(2)}$V$_{15.4(5)}$Zr$_{24.3(1)}$, which is in agreement with the starting composition.

\begin{figure}
\begin{center}
\includegraphics[width=0.8\linewidth]{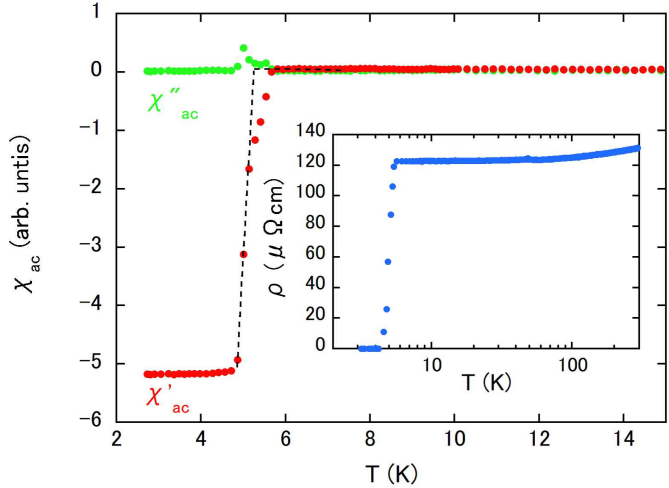}
\end{center}
\caption{Temperature dependences of $\chi_{ac}$ and $\rho$ of Hf$_{21}$Nb$_{25}$Ti$_{15}$V$_{15}$Zr$_{24}$.}
\label{f2}
\end{figure}

Shown in Fig.\ 2 is $\chi_{ac}$ (T) of Hf$_{21}$Nb$_{25}$Ti$_{15}$V$_{15}$Zr$_{24}$.
The real part of $\chi_{ac}$, $\chi_{ac}^{'}$, exhibits the diamagnetic signal below the superconducting critical temperature $T_\mathrm{c}$=5.3 K, which is determined as being the intercept of the linearly extrapolated diamagnetic slope with the normal state signal\cite{Rohr:PNAS2016,Hamamoto:MRX2018} (see the broken lines in Fig.\ 2).
The imaginary $\chi_{ac}$, $\chi_{ac}^{"}$, shows a peak, which is also characteristic of superconductors.
In the inset of Fig.\ 2, $\rho$ (T) exhibits a very weak temperature dependence, reflecting the atomic disorder.
$\rho$ rapidly drops below approximately $T_\mathrm{c}$ and shows the zero resistivity.

\begin{table}
\caption{Comparison of cocktail effect of $T_\mathrm{c}$, $S_\mathrm{mix}$/$R$ and valence electron concentration (VEC) per atom among BCC HEA superconductors. $T_\mathrm{c}^\mathrm{obs.}$ and $T_\mathrm{c}^\mathrm{base}$ are the experimental $T_\mathrm{c}$ and $T_\mathrm{c}$ obtained by averaging $T_\mathrm{c}$'s of the constituent elements weighted by each atomic percentage, respectively.}
\label{t1}
\begin{tabular}{ccccccc}
\hline
\scriptsize{Compound} & \scriptsize{$T_\mathrm{c}^\mathrm{obs.}$ (K)} & \scriptsize{$T_\mathrm{c}^\mathrm{base}$ (K)} & \scriptsize{$T_\mathrm{c}^\mathrm{obs.}/T_\mathrm{c}^\mathrm{base}$} & \scriptsize{$S_\mathrm{mix}$/$R$} & \scriptsize{VEC} & \scriptsize{ref.} \\
\hline
\scriptsize{Hf$_{21}$Nb$_{25}$Ti$_{15}$V$_{15}$Zr$_{24}$} & \scriptsize{5.3} & \scriptsize{3.30} & \scriptsize{1.61} & \scriptsize{1.586} & \scriptsize{4.4} & \scriptsize{this work} \\
\scriptsize{Ta$_{34}$Nb$_{33}$Hf$_{8}$Zr$_{14}$Ti$_{11}$} & \scriptsize{7.3} & \scriptsize{4.72} & \scriptsize{1.55} & \scriptsize{1.453} & \scriptsize{4.67} & \scriptsize{\cite{Kozelj:PRL2014}} \\
\scriptsize{Nb$_{20}$Re$_{20}$Hf$_{20}$Zr$_{20}$Ti$_{20}$} & \scriptsize{5.3} & \scriptsize{2.42} & \scriptsize{2.2} & \scriptsize{1.609} & \scriptsize{4.8} & \scriptsize{\cite{Marik:JALCOM2018}} \\
\scriptsize{(TaNb)$_{0.7}$(ZrHfTi)$_{0.3}$} & \scriptsize{8.0} & \scriptsize{4.93} & \scriptsize{1.62} & \scriptsize{1.195} & \scriptsize{4.7} & \scriptsize{\cite{Rohr:PNAS2016}} \\
\scriptsize{(TaNb)$_{0.6}$(ZrHfTi)$_{0.4}$} & \scriptsize{7.56} & \scriptsize{4.28} & \scriptsize{1.77} & \scriptsize{1.528} & \scriptsize{4.6} & \scriptsize{\cite{Rohr:PNAS2016}} \\
\scriptsize{(TaNb)$_{0.5}$(ZrHfTi)$_{0.5}$} & \scriptsize{6.46} & \scriptsize{3.63} & \scriptsize{1.78} & \scriptsize{1.589} & \scriptsize{4.5} & \scriptsize{\cite{Rohr:PNAS2016}} \\
\scriptsize{(TaNb)$_{0.16}$(ZrHfTi)$_{0.84}$} & \scriptsize{4.52} & \scriptsize{1.41} & \scriptsize{3.21} & \scriptsize{1.473} & \scriptsize{4.16} & \scriptsize{\cite{Rohr:PNAS2016}} \\
\scriptsize{Nb$_{66}$Ti$_{33}$} & \scriptsize{9.2} & \scriptsize{6.34} & \scriptsize{1.45} & \scriptsize{0.6365} & \scriptsize{4.55} & \scriptsize{\cite{Rohr:PNAS2016}} \\
\hline
\end{tabular}
\end{table}

\begin{figure}
\begin{center}
\includegraphics[width=0.8\linewidth]{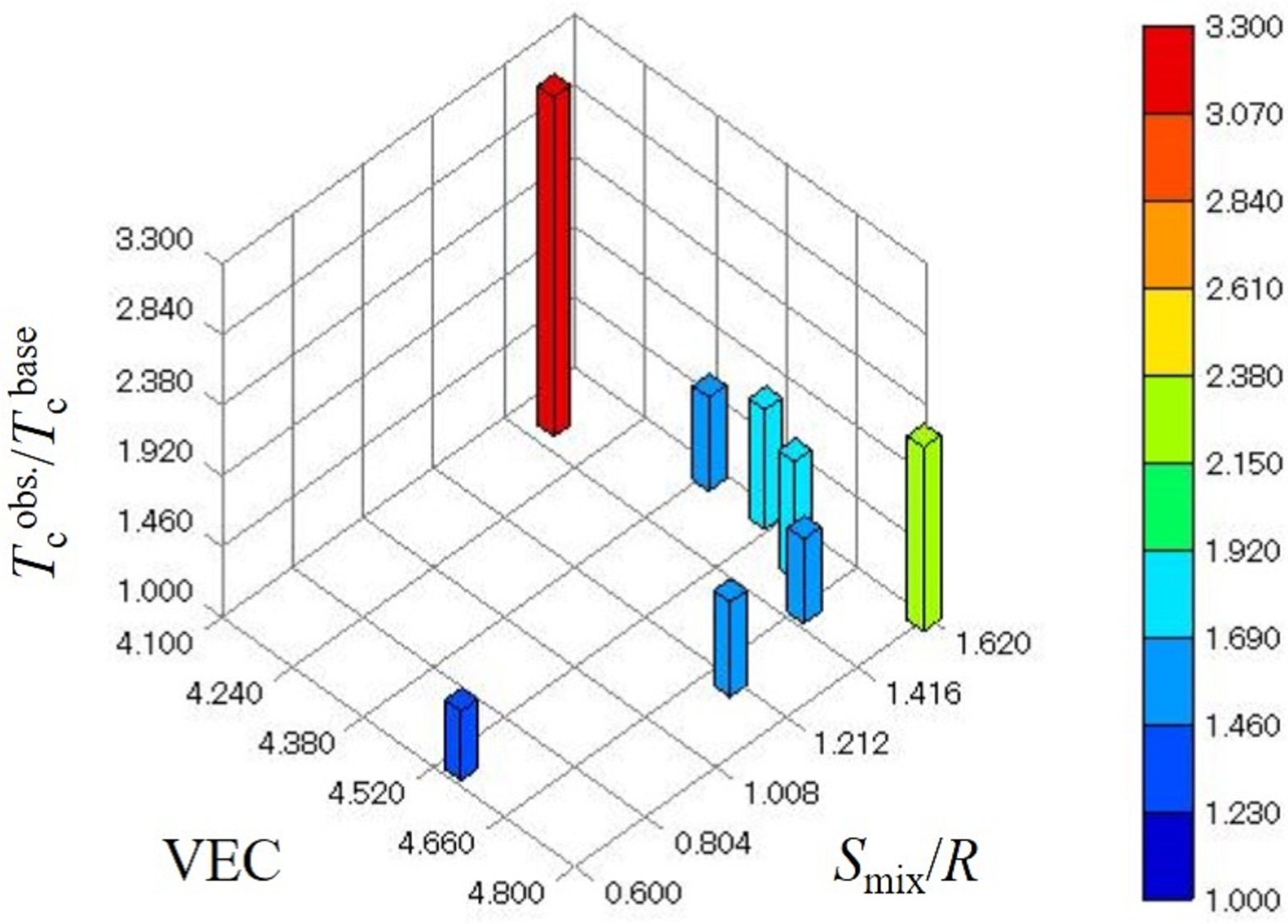}
\end{center}
\caption{3D bar graph of $T_\mathrm{c}^\mathrm{obs.}/T_\mathrm{c}^\mathrm{base}$ as a function of $S_\mathrm{mix}$/$R$ and VEC.}
\label{f3}
\end{figure}

One of the interesting properties of HEAs is the cocktail effect, which means an enhancement of physical property beyond the simple mixture of those of constituent elements.
Table 1 shows the comparison of cocktail effect of $T_\mathrm{c}$, $S_\mathrm{mix}$/$R$ and valence electron concentration (VEC) per atom of BCC HEA superconductors.
We also list a reference material Nb$_{66}$Ti$_{33}$, that does not fulfill both of two HEA-definitions mentioned above.
Following the method reported in ref.\cite{Kozelj:PRL2014}, the cocktail effect of $T_\mathrm{c}$ is evaluated by $T_\mathrm{c}^\mathrm{obs.}/T_\mathrm{c}^\mathrm{base}$, where $T_\mathrm{c}^\mathrm{obs.}$ and $T_\mathrm{c}^\mathrm{base}$ are the experimental $T_\mathrm{c}$ and $T_\mathrm{c}$ obtained by averaging $T_\mathrm{c}$'s of the constituent elements weighted by each atomic percentage, respectively.
$T_\mathrm{c}^\mathrm{base}$ plays a role of base line in order to know an enhancement of $T_\mathrm{c}$ due to a mixing of constituent elements, and a larger $T_\mathrm{c}^\mathrm{obs.}/T_\mathrm{c}^\mathrm{base}$ indicates a strong cocktail effect.
VEC is an important factor deciding $T_\mathrm{c}$, that is well known as the Matthias rule\cite{Matthias:PR1955}, in which the VEC dependence of $T_\mathrm{c}$ for crystalline $d$-electron systems shows a maximum at approximately VEC=4.5 and $T_\mathrm{c}$ steeply decreases to zero around VEC=4.0 or 5.3.
However, the BCC HEA superconductors\cite{Rohr:PNAS2016} preserve a rather high $T_\mathrm{c}$ compared to the prediction value of the Matthias rule especially at the side of VEC smaller than 4.5.
Keeping this in mind, we made a three-dimensional bar graph in Fig.\ 3 by using the numerical values listed in Table 1.
The enhancement of $T_\mathrm{c}$ in each HEA is stronger than that of Nb$_{66}$Ti$_{33}$.
The value of $T_\mathrm{c}^\mathrm{obs.}/T_\mathrm{c}^\mathrm{base}$ seems to increase with increasing $S_\mathrm{mix}$.
We note here that (TaNb)$_{0.16}$(ZrHfTi)$_{0.84}$ shows the strong $T_\mathrm{c}$ enhancement, although $S_\mathrm{mix}$/$R$ of this compound is relatively small.
VEC of (TaNb)$_{0.16}$(ZrHfTi)$_{0.84}$ is 4.16, where $T_\mathrm{c}$ is suppressed to nearly zero in the VEC dependence of $T_\mathrm{c}$ by the Matthias rule.
This result suggests that the VEC also somewhat affects the enhancement of $T_\mathrm{c}$ in HEA superconductors.

\section{Summary}
We have found that Hf$_{21}$Nb$_{25}$Ti$_{15}$V$_{15}$Zr$_{24}$ is a new member of HEA superconductors. It crystallizes into the BCC structure with $a$=3.401 \AA, and $T_\mathrm{c}$ is 5.3 K. The cocktail effect of $T_\mathrm{c}$ in BCC HEA superconductors might be affected by $S_\mathrm{mix}$ and/or VEC.

\section*{Acknowledgments}
J.K. is grateful for the support provided by Comprehensive Research Organization of Fukuoka Institute of Technology.

\end{document}